\documentclass[aps,pra,showpacs,preprint]{revtex4}
\usepackage{graphicx}
\usepackage{amsmath,amssymb,amsfonts}
\usepackage{dcolumn}
\usepackage[english]{babel}
\usepackage{epsfig}

\begin{document}

\title{Parabolic Sturmians approach to the three-body continuum Coulomb problem}
\author{S.~A.~Zaytsev$^1$},
\email[E-mail: ]{zaytsev@fizika.khstu.ru}
\author{Yu.~V.~Popov$^2$, B.~Piraux$^3$}
\affiliation{$^1$Pacific National University, Khabarovsk, 680035, Russia\\
             $^2$Nuclear Physics Institute, Moscow State University, Moscow, 119991, Russia\\
             $^3$Institute of Condensed Matter and Nanosciences, Universit\'e catholique de Louvain,\\
             B\^{a}timent de Hemptinne, 2, chemin du cyclotron, B1348 Louvain-la-Neuve, Belgium}

\begin{abstract}

The three-body continuum Coulomb problem is treated in terms of the
generalized parabolic coordinates. Approximate solutions are
expressed in the form of a Lippmann-Schwinger type equation, where the
Green's function includes the leading term of the kinetic energy and
the total potential energy, whereas the potential contains the
non-orthogonal part of the kinetic energy operator. As a test of
this approach, the integral equation for the $(e^-,\, e^-,\,
{\mbox{He}^{++}})$ system is solved numerically by using the
parabolic Sturmian basis representation of the (approximate)
potential. Convergence of the expansion coefficients of the solution
is obtained as the basis set used to describe the potential is
enlarged.

\end{abstract}
\maketitle

\section{Introduction}
The three-body continuum Coulomb problem is one of the fundamental
unresolved problems of theoretical physics. In atomic physics, a
prototype example is a two-electron continuum which arises as a
final state in electron-impact ionization and double photoionization
of atomic systems. Several discrete-basis-set methods for the
calculation of such processes have recently been developed including
convergent close coupling (CCC) \cite{CCC1,CCC2}, the
Coulomb-Sturmian separable expansion method \cite{TPF1,TPF2}, the
J-matrix method \cite{JM1,JM2,JM3}. In all these approaches (see
also \cite{Sturm1}) the continuous Hamiltonian spectrum is
represented in the context of complete square integrable bases.
Despite the enormous progress made so far in discretization and
subsequent numerical solutions of three-body differential and
integral equations of the Coulomb scattering theory, a number of
related mathematical problems remain open. Actually, the use of a
product of two fixed charge Coulomb waves for the two outgoing
electrons as an approximation to the three-body continuum state is
typical of these approaches. As a consequence, a long-range
potential appears in the kernel of the corresponding
Lippmann-Schwinger equation. Since this integral equation is
non-compact, its solution therefore is divergent as the size of a
$L^2$ basis used to describe the potential is increased. Note that a
renormalization approach can not cure this problem.

A theoretical treatment of Coulomb breakup problems (which does not
require screening or any regularization) has been recently suggested
by Kadyrov et al. \cite{Kadyrov8}.

In several papers \cite{JP1,JP2,JP3} a new approach for solution of
the three-body continuum Coulomb problem was introduced. The
development of the method is chiefly based upon the fact that the
asymptotic wave operator, which determines the wave function
behavior when all interparticle distances are large (in the
$\Omega_0$ domain), is separable in terms of generalized parabolic
coordinates \cite{Klar}. The parabolic coordinate eigenfunctions of
the asymptotic wave operator which satisfy the Redmond's conditions
\cite{Redmond} in the asymptotic domain $\Omega_0$ are expressed in
terms of a product of three two-body Coulomb functions, each
depending of one parabolic coordinate. These functions, often called
the C3 (or BBK) wave functions \cite{C31,C32,C33}, are successfully
used as the final-state wave functions for calculating cross
sections for electron-impact ionization and double photoionization
of helium \cite{Jones2003,Ancarani2004,Ch10}. In \cite{JP2} it has
been proposed to use an integral equation of a Lippmann-Schwinger
type to construct an approximate solution that describe three
charged particles moving in the three-body continuum. In this
integral equation the Green's function in the whole configuration
space (and not just in the domain $\Omega_0$) is approximated by the
inverse of the asymptotic wave operator. In turn, the well-known
non-orthogonal part of the kinetic energy operator, which represents
the difference between the total and asymptotic wave operators,
plays the role of the potential. Asymptotic behavior of solutions is
determined by the inhomogeneous term which is given by the C3 wave
function.

To test the practicality of the Lippmann-Schwinger equation
approach, we restrict ourselves to the so-called outgoing
approximation \cite{Gasaneo97}, which assumes that the sought-for
solution (as well as the inhomogeneity) depends  only on the
parabolic coordinates $\xi_j$, $j=1,\,2,\,3$, and therefore the
approximate potential operator contains only the terms which involve
the mixed derivatives $\partial^2/\partial \xi_l
\partial \xi_s$, $l\neq s$. In this work we address the question
of compactness of the kernel of the integral equation. Any compact
operator may be approximated arbitrary closely by an operator of
finite rank. In order to study the properties of the kernel we
choose the parabolic Sturmian basis set \cite{Ojha1} and construct a
sequence of separable kernels. Then to check the existence of a
limit of this sequence we examine the convergence behavior of the
first few expansion coefficients of the solution as the basis set is
increased. The coefficients are found to exhibit oscillations whose
amplitude does not decrease as the number of terms in the
representation of the potential grows. This result is similar to the
Gibbs' phenomenon known from Fourier analysis (see, e. g.,
\cite{Stone,Canuto}), where the oscillation of an approximant about
the exact function (which possesses a discontinuity) is a
consequence of the abrupt truncation of the Fourier sum.  In order
to avoid or at least to reduce the Gibbs' phenomenon, smoothing
procedures are used that attenuate the higher order coefficients
\cite{Stone,Canuto}. In this paper, we use the Lanczos smoothing
factors, introduced in the potential separable expansion (PSE)
method \cite{PSE} (see also \cite{Papp0} and references therein), in
constructing the basis set representations of the potential.

This paper is organized as follows. In Sec.~II we introduce the
notations, recall the generalized parabolic coordinates definition
and express a formal solution for the three-body Coulomb problem in
the form of the Lippmann-Schwinger-type equation. In Sec.~III we
briefly outline the parabolic Sturmians approach. In particular, we
present the matrix representation of the three-body Coulomb Green
function and consider the potential operator approximation. In
Sec.~IV calculations of the continuum state of the $(e^-,\, e^-,
\,\mbox{He}^{++})$ system, where both electrons recede from the
residual ion in opposite directions with equal energies, are
described. Our aim is to study the rate of convergence as the basis
set used to describe the potential operator is enlarged. If the same
number of basis functions for each parabolic coordinate is used, the
problem of numerical solution rapidly gets out of hand. Thus, in the
separable expansion of the potential, the number of basis functions
for the three chosen curvilinear parabolic coordinates $\xi_1,\,
\xi_2,\, \xi_3$ is increased, whereas a single basis function is
taken for each of the remaining three coordinates $\eta_1,\,
\eta_2,\, \eta_3$. The calculations show that the convergence on a
basis of reasonable size can be obtained by using the Lanczos
smoothing factors. Sec.~V contains a brief discussion of the overall
results. Atomic units are used throughout.

\section{Theory}
We consider three particles of masses $m_1$, $m_1$, $m_3$, charges
$Z_1$, $Z_2$, $Z_3$ and momenta ${\bf k}_1$, ${\bf k}_2$, ${\bf
k}_3$. The Hamiltonian of the system in the center of mass frame is
given by
\begin{equation}\label{HS}
    \hat{H}=-\frac{1}{2\mu_{12}}\Delta_{\bf R}
    -\frac{1}{2\mu_{3}}\Delta_{\bf r}+\frac{Z_1Z_2}{r_{12}}
    +\frac{Z_2Z_3}{r_{23}}+\frac{Z_1Z_3}{r_{13}},
\end{equation}
where ${\bf r}_{ls}$ denotes the relative coordinates
\begin{equation}\label{rc}
{\bf r}_{ls}={\bf r}_l-{\bf r}_s, \quad r_{ls}=\left|{\bf r}_{ls}
\right|,
\end{equation}
${\bf R}$ and ${\bf r}$ are the Jacobi
coordinates
\begin{equation}\label{Rr} {\bf R}={\bf r}_1-{\bf r}_2,
\quad {\bf r}={\bf r}_3-\frac{m_1{\bf r}_1+m_2{\bf r}_2}{m_1+m_2}.
\end{equation}
The reduced masses are defined as
\begin{equation}\label{mass}
    \mu_{12}=\frac{m_1m_2}{m_1+m_2}, \quad
    \mu_{3}=\frac{m_3\left(m_1+m_2\right)}{m_1+m_2+m_3}.
\end{equation}
In the Schr\"{o}dinger equation
\begin{equation}\label{SE}
    \hat{H}\Phi=E\Phi
\end{equation}
the eigenenergy $E>0$ is given by
\begin{equation}\label{Eng}
    E=\frac{1}{2\mu_{12}}\,{\bf
K}^2+\frac{1}{2\mu_{3}}\,{\bf k}^2,
\end{equation}
where ${\bf K}$ and ${\bf k}$ are the momenta conjugate to the
variables ${\bf R}$ and ${\bf r}$. Substituting
\begin{equation}\label{PsiO}
    \Phi=e^{i({\bf K}\cdot{\bf R}+{\bf k}\cdot{\bf r})}\Psi
\end{equation}
into (\ref{SE}), we arrive at the equation for the reduced wave
function $\Psi$
\begin{equation}\label{SEO}
    \left[-\frac{1}{2\mu_{12}}\,\Delta_{\bf R}
    -\frac{1}{2\mu_{3}}\,\Delta_{\bf r}
    -\frac{i}{\mu_{12}}\,{\bf K}\cdot \nabla_{\bf R}
    -\frac{i}{\mu_{3}}\,{\bf k}\cdot \nabla_{\bf r}
    +\frac{Z_1Z_2}{r_{12}}+\frac{Z_2Z_3}{r_{23}}
    +\frac{Z_1Z_3}{r_{13}}\right]\Psi=0.
\end{equation}
Leading-order asymptotic terms of $\Psi$ in the $\Omega_0$ domain
are expressed in terms of the generalized parabolic coordinates
\cite{Klar}
\begin{equation}\label{pc}
  \begin{array}{c}
  \xi_1=r_{23}+\hat{\bf k}_{23}\cdot{\bf r}_{23}, \quad
  \eta_1=r_{23}-\hat{\bf k}_{23}\cdot{\bf r}_{23},\\
  \xi_2=r_{13}+\hat{\bf k}_{13}\cdot{\bf r}_{13}, \quad
  \eta_2=r_{13}-\hat{\bf k}_{13}\cdot{\bf r}_{13},\\
  \xi_3=r_{12}+\hat{\bf k}_{12}\cdot{\bf r}_{12}, \quad
  \eta_3=r_{12}-\hat{\bf k}_{12}\cdot{\bf r}_{12},\\
  \end{array}
\end{equation}
where ${\bf k}_{ls}=\frac{{\bf k}_l m_s-{\bf k}_s m_l}{m_l+m_s}$ is
the relative momentum, $\hat{\bf k}_{ls}=\frac{{\bf
k}_{ls}}{k_{ls}}$, $k_{ls}=\left|{\bf k}_{ls}\right|$. The operator
in the square brackets, denoted by $\hat{D}$, can be decomposed into
two terms \cite{Klar}
\begin{equation}\label{PSEO}
    \hat{D}=\hat{D}_0+\hat{D}_1,
\end{equation}
where the operator $\hat{D}_0$ contains the leading term of the
kinetic energy and the total potential energy:
\begin{equation}\label{D0}
 \begin{array}{c}
    \hat{D}_0=\sum \limits_{j=1}^{3}\frac{1}{\mu_{ls}\left(\xi_j+\eta_j \right)}
    \left[ \hat{h}_{\xi_j}+\hat{h}_{\eta_j}+2k_{ls}
    t_{ls}\right],\\[3mm]
    \mbox{for } j\neq l,\,s \mbox{ and } l<s,\\
 \end{array}
\end{equation}
\begin{equation}\label{hxi}
    \hat{h}_{\xi_j}=-2\left(\frac{\partial}{\partial \xi_j}
    \xi_j\frac{\partial}{\partial \xi_j}+i k_{ls}\xi_j\frac{\partial}{\partial \xi_j} \right),
\end{equation}
\begin{equation}\label{heta}
    \hat{h}_{\eta_j}=-2\left(\frac{\partial}{\partial \eta_j}
    \eta_j\frac{\partial}{\partial \eta_j}-i k_{ls}\eta_j\frac{\partial}{\partial \eta_j}
    \right).
\end{equation}
Here $t_{ls}=\frac{Z_l Z_s \mu_{ls}}{k_{ls}}$, $\mu_{ls}=\frac{m_l
m_s}{m_l+m_s}$. The operator $\hat{D}_1$ represents the remaining
part of the kinetic energy \cite{Klar} which in the case of the
$(e^-, \, e^-, \, \mbox{He}^{++})=(123)$ system with $m_3=\infty$
takes the form \cite{Gasaneo97}
\begin{equation}\label{D1}
 \begin{array}{c}
 \hat{D}_1=\sum \limits _{j=1}^{2}(-1)^{j+1}\left[{\bf u}_j^{-}\cdot{\bf u}_3^{-}
 \frac{\partial^2}{\partial \xi_j\, \partial \xi_3}+
 {\bf u}_j^{-}\cdot{\bf u}_3^{+}
 \frac{\partial^2}{\partial \xi_j\, \partial \eta_3} \right.\\[4mm]
+\left.{\bf u}_j^{+}\cdot{\bf u}_3^{-}
 \frac{\partial^2}{\partial \eta_j\, \partial \xi_3}+
 {\bf u}_j^{+}\cdot{\bf u}_3^{+}
 \frac{\partial^2}{\partial \eta_j\, \partial \eta_3} \right],\\
 \end{array}
\end{equation}
where
\begin{equation}\label{uvec}
    {\bf u}_j^{\pm}=\hat{\bf r}_{ls}\mp \hat{\bf k}_{ls}.
\end{equation}

The asymptotic behavior of solutions $\Psi$ is determined by the
operator $\hat{D}_0$. In particular, there exist solutions to the
equation
\begin{equation}\label{C3E}
 \hat{D}_0 \Psi_{C3}=0,
\end{equation}
which satisfy the Redmond conditions in $\Omega_0$. These solutions
are well-known the C3 wave functions. $\Psi_{C3}$ is expressed in
terms of a product of three Coulomb waves. For example, $\Psi_{C3}$
with pure outgoing behavior is written as
\begin{equation}\label{PsiC3}
    \Psi_{C3} = \prod \limits _{j=1}^{3}
    {_1F_1}\left(i t_{ls},\, 1;\; -i k_{ls}\xi_j \right).
\end{equation}
In turn, $\hat{D}_1$ is regarded as a perturbation which does not
violate the asymptotic conditions \cite{Klar}.

Our goal is to construct an approximate solution $\Psi$ of
(\ref{SEO}) that satisfies the boundary condition (\ref{PsiC3}) in
the asymptotic $\Omega_0$ domain. For this purpose, we rewrite
(\ref{SEO}) in terms of the operators
\begin{equation}\label{h}
 \begin{array}{c}
\hat{\mathcal{H}}\equiv\prod \limits
_{j=1}^3\mu_{ls}\left(\xi_j+\eta_j
\right)\hat{D}_0=\mu_{13}\left(\xi_2+\eta_2
\right)\mu_{12}\left(\xi_3+\eta_3 \right)\hat{\mathfrak{h}}_1\\[3mm]
+\mu_{23}\left(\xi_1+\eta_1 \right)\mu_{12}\left(\xi_3+\eta_3
\right)\hat{\mathfrak{h}}_2 +\mu_{23}\left(\xi_1+\eta_1
\right)\mu_{13}\left(\xi_2+\eta_2 \right)\hat{\mathfrak{h}}_3,\\
 \end{array}
\end{equation}
\begin{equation}\label{hj}
\hat{\mathfrak{h}}_j=\hat{h}_{\xi_j}+\hat{h}_{\eta_j}+2k_{ls}t_{ls},
\end{equation}
and
\begin{equation}\label{V}
\hat{\mathcal{V}}\equiv\prod \limits
_{j=1}^3\mu_{ls}\left(\xi_j+\eta_j \right)\hat{D}_1
\end{equation}
after multiplying on the left by $\prod \limits
_{j=1}^3\mu_{ls}\left(\xi_j+\eta_j \right)$:
\begin{equation}\label{PSEM}
    \left[\hat{\mathcal{H}}+ \hat{\mathcal{V}}\right]\Psi=0.
\end{equation}
Thus, given the Green's function operator
$\hat{\mathcal{G}}=\hat{\mathcal{H}}^{-1}$, we can take into account
the non-orthogonal term $\hat{D}_1$ of the kinetic energy operator
(which is larger than the total potential in the ``inner zone''
\cite{Berakdar}) by putting it into the kernel of the
Lippmann-Schwinger type equation:
\begin{equation}\label{LSE}
    \Psi=\Psi_{C3}-\hat{\mathcal{G}}\hat{\mathcal{V}}\Psi.
\end{equation}

\subsection*{Green's functions}

Based on the fact that the original operator $\hat{D}_0$ is
separable in the parabolic coordinates (\ref{pc}), the inverse of
the six-dimensional operator $\hat{\mathcal{H}}$ (\ref{h}) can be
expressed as a convolution of the three two-dimensional Green's
function operators $\hat{G}_j^{(\pm)}$ whose kernels
\begin{equation}\label{G2K}
    \left< \xi_j,\, \eta_j\right|\hat{G}_j^{(\pm)}\left(t_{ls},\,\mathcal{E}_j\right) \left|\xi'_j,\, \eta'_j \right>
    \equiv G_j^{(\pm)}\left(t_{ls},\,\mathcal{E}_j;
\; \xi_j,\, \eta_j, \, \xi'_j,\, \eta'_j \right)
\end{equation}
satisfy the equations
\begin{equation}\label{TwoD}
\left[\hat{\mathfrak{h}}_j+\mu_{ls}\,C_j\left(\xi_j+\eta_j\right)\right]G_j^{(\pm)}\left(t_{ls},\,\mathcal{E}_j;
\; \xi_j,\, \eta_j, \, \xi'_j,\, \eta'_j \right)=
\delta\left(\xi_j-\xi'_j\right)\delta\left(\eta_j-\eta'_j\right).
\end{equation}
In view of (\ref{h}) the separation parameters $C_j$ are subject to
the condition
\begin{equation}\label{Cj}
    C_1+C_2+C_3=0.
\end{equation}
In (\ref{G2K}) and (\ref{TwoD}) we introduced auxiliary variables:
\begin{equation}\label{Egma}
    \mathcal{E}_j=\frac{\gamma_j^2}{2}=\frac{k_{ls}^2}{2}-\mu_{ls}\,C_j.
\end{equation}
The technique presented in \cite{Swainson} can be employed to derive
useful forms for the two-dimensional Green's function. For example,
$G_j^{(\pm)}$ can be expressed in the form (for simplicity we omit
the indices):
\begin{equation}\label{G22}
 \begin{array}{c}
G^{(\pm)}\left(t,\, \mathcal{E}; \; \xi,\, \eta, \, \xi',\, \eta'
\right)=\mp\frac{i
\gamma}{4}\,e^{\frac{i}{2}k(\xi'-\xi+\eta-\eta')}\,\int
\limits_{0}^{\infty} dz\,\sinh(z)\,
\left[\coth\left(\frac{z}{2}\right) \right]^{\mp 2 i
     \tau}\\[3mm]
     \times e^{\pm
     i\frac{\gamma}{2}(\xi+\xi'+\eta+\eta')\,\cosh(z)}\,I_0\left( \mp i \gamma \sqrt{\xi\, \xi'}
     \sinh(z)\right)\,I_0\left( \mp i \gamma \sqrt{\eta\, \eta'}
     \sinh(z)\right),\\
 \end{array}
\end{equation}
where
\begin{equation}\label{tau}
    \tau = \frac{k}{\gamma}\,t,
\end{equation}
$I_{\nu}(z)$ is the modified Bessel function of order $\nu$
\cite{Abramowitz}.

The resolvent of $\hat{\mathcal{H}}$ standard representation
\cite{FMB} involves integration along two contours which encircle
the spectra of two of the wave operators $\hat{\mathfrak{h}}_j$,
$j=1,\, 2,\,3$. In the paper \cite{JP3} an integral representation
of $\hat{\mathcal{H}}^{-1}$ suitable for numerical computation has
been proposed (see below).

\section{Parabolic Sturmians approach}
If the kernel $\hat{\mathcal{G}}\hat{\mathcal{V}}$ is compact, then
the integral equation (\ref{LSE}) can be solved by, e. g., the
algebraic method in which the potential $\hat{\mathcal{V}}$ is
approximated by operators of finite rank. For the expansion of
$\hat{\mathcal{V}}$ we use a set of square-integrable parabolic
Sturmian functions \cite{Ojha1}
\begin{equation}\label{B6}
    \left|\mathfrak{N} \right>=\prod \limits _{j=1}^{3} \phi_{n_j\,m_j}\left(\xi_j,\,\eta_j
    \right),
\end{equation}
\begin{equation}\label{B2}
\phi_{n_j\,m_j}\left(\xi_j,\,\eta_j \right)=\psi_{n_j}\left(\xi_j
\right)\psi_{m_j}\left(\eta_j\right),
\end{equation}
\begin{equation}\label{B1}
    \psi_{n}\left(x\right)=\sqrt{2b_j}\,e^{-b_j x}L_{n}(2b_j x).
\end{equation}
The basis functions (\ref{B2}), (\ref{B1}) are parametrized with a
separate Sturmian exponent $b_j$ for each pair $\left\{\xi_j,\,
\eta_j\right\}$, $j=\overline{1,\,3}$. Thus, the operator
$\hat{\mathcal{V}}$ is represented by its projection
$\hat{\mathcal{V}}^{\mathcal{N}}$ onto a subspace of basis
functions,
\begin{equation}\label{VN}
\hat{\mathcal{V}}^{\mathcal{N}}= \sum \limits _{j=1}^3 \, \sum
\limits _{n_j,\, n_j'=0}^{N_j-1}\,\sum \limits _{m_j,\,
m_j'=0}^{M_j-1}
\left|\mathfrak{N}\right>\left<\mathfrak{N}\right|\hat{\mathcal{V}}
\left|\mathfrak{N}'\right>\left<\mathfrak{N}'\right|,
\end{equation}
and the solution $\Psi$ of the problem is obtained for
$\hat{\mathcal{V}}^{\mathcal{N}}$. Inserting
$\hat{\mathcal{V}}^{\mathcal{N}}$ into Eq. (\ref{LSE}) then leads to
a finite matrix equation for the expansion coefficients
$\left[\underline{a}\right]_{\mathfrak{N}}=\left<\mathfrak{N}\right|\Psi\left.
\right>$,
\begin{equation}\label{DALSE}
    \underline{a}=\underline{a}^{(0)}-\underline{\mathcal{G}}\,\underline{\mathcal{V}}\,\underline{a},
\end{equation}
which has the solution
\begin{equation}\label{DALSES}
    \underline{a}=\left(\underline{1}+\underline{\mathcal{G}}\,\underline{\mathcal{V}} \right)^{-1}
    \underline{a}^{(0)}.
\end{equation}
Here
$\left[\underline{\mathcal{G}}\right]_{\mathfrak{N}\mathfrak{N}'}=\left<\mathfrak{N}\right|\hat{\mathcal{G}}
\left|\mathfrak{N}'\right>$ and
$\left[\underline{\mathcal{V}}\right]_{\mathfrak{N}\mathfrak{N}'}=\left<\mathfrak{N}\right|\hat{\mathcal{V}}
\left|\mathfrak{N}'\right>$ are the Green's function operator and
potential operator matrices of order $\mathcal{N}=\prod_{j=1}^3
N_j\,M_j$, and $\underline{a}^{(0)}$ is the coefficient vector of
$\Psi_{C3}$. The wave function $\Psi$ is expressed in terms of the
solution of Eq. (\ref{DALSE}):
\begin{equation}\label{PsiN}
    \Psi = \Psi_{C3}-\sum \limits _{j=1}^3 \, \sum
\limits _{n_j=0}^{N_j-1}\,\sum \limits
_{m_j=0}^{M_j-1}\left[\underline{C}\right]_{\mathfrak{N}}\hat{\mathcal{G}}\left|\mathfrak{N}\right>,
\end{equation}
where $\underline{C}=\underline{\mathcal{V}}\,\underline{a}$.

\subsection*{Green's function matrices}
To construct the six-dimensional Green's function matrix
$\underline{\mathcal{G}}$, we need the two-dimensional Green's
function (\ref{G22}) matrix with elements
\begin{equation}\label{Gnmn_m_}
 \begin{array}{c}
 G^{(\pm)}_{n,\,m;\;n',\,m'}\left(t,\, \mathcal{E}\right)
 =\int\limits_{0}^{\infty}\int\limits_{0}^{\infty}
 \int\limits_{0}^{\infty}\int\limits_{0}^{\infty}d\xi\,d\eta\,d\xi'\,d\eta'
 \left<\phi_{n\,m}\right|\left.\xi,\, \eta\right>\\
 \times\left<\xi,\, \eta \right|\hat{G}^{(\pm)}\left(t,\, \mathcal{E}
 \right)\left|\xi',\, \eta' \right>\left<\xi',\, \eta' \right.\left|\phi_{n'\,m'}\right>.\\
 \end{array}
\end{equation}
Inserting (\ref{G22}) into (\ref{Gnmn_m_}), we obtain after some
simple but tedious algebraic manipulations
\begin{equation}\label{Gnmn_mp}
 \begin{array}{c}
 G^{(+)}_{n,\,m;\;n',\,m'}\left(t,\, \mathcal{E}\right)
 =\frac{i}{2\gamma}\left(\frac{\zeta-1}{\zeta} \right)\,
  \frac{(-\theta)^{n+m'}}{(-\lambda)^{n'+m}}\sum\limits_{\ell=0}^{\nu+\mu}\,c_{\ell}\,\zeta^{\ell}\;
 \frac{\Gamma\left(i \tau+\ell+1 \right)\,\Gamma\left(K+1-2\ell\right)}
 {\Gamma\left(i \tau+K+2-\ell\right)}\\[4mm]
 \times{_2F_1\left(K+1-2\ell,\, i \tau-\ell;\;
 i\tau+K+2-\ell;\; \zeta^{-1}\right)},\\
 \end{array}
\end{equation}
where
\begin{equation}\label{tlz}
    K = n+n'+m+m',\quad \theta= \frac{2b+i(\gamma-k)}{2b-i(\gamma-k)}\,,
    \quad \lambda= \frac{2b-i(\gamma+k)}{2b+i(\gamma+k)}\,,
    \quad \zeta=\frac{\lambda}{\theta}\,,
\end{equation}

\begin{equation}\label{cl}
    c_{\ell}= \sum \limits _{j=\max(\ell-\mu,\, 0)}^{\min(\ell, \,
    \nu)} {n\choose j}{n'\choose j}{m\choose \ell-j}{m'\choose
    \ell-j}, \quad \nu=\min(n,\,n'), \quad \mu=\min(m,\,m').
\end{equation}

Replacing $\gamma$ by $-\gamma$ ($\lambda \rightarrow \theta$,
$\theta \rightarrow \lambda$, $\zeta \rightarrow 1/\zeta$, $\tau
\rightarrow -\tau$) in (\ref{Gnmn_mp}) gives
\begin{equation}\label{Gnmn_mm}
 \begin{array}{c}
 G^{(-)}_{n,\,m;\;n',\,m'}\left(t,\, \mathcal{E}\right)
 =\frac{i}{2\gamma}(\zeta-1)\,
  \frac{(-\lambda)^{n+m'}}{(-\theta)^{n'+m}}\sum\limits_{\ell=0}^{\nu+\mu}\,c_{\ell}\,\zeta^{-\ell}\;
 \frac{\Gamma\left(-i \tau+\ell+1 \right)\,\Gamma\left(K+1-2\ell\right)}
 {\Gamma\left(-i \tau+K+2-\ell\right)}\\[4mm]
 \times{_2F_1\left(K+1-2\ell,\,-i \tau-\ell;\;
 -i\tau+K+2-\ell;\; \zeta\right)}.\\
 \end{array}
\end{equation}

\begin{figure*}[ht]
\centerline{\psfig{figure=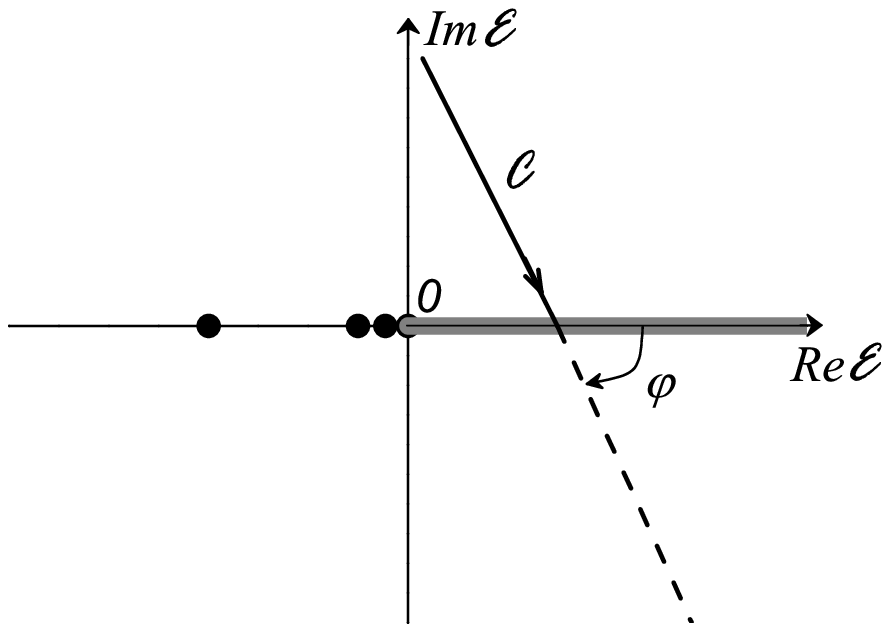,width=1\textwidth}}
\caption{The path of integration $\mathcal{C}$. The solid line is
the part of $\mathcal{C}$ which remains on the physical sheet. The
part of $\mathcal{C}$ which moves onto the unphysical sheet is shown
by the dashed line.} \label{fig1}
\end{figure*}

The elements $G^{(\pm)}_{n_j,\,m_j;\;n'_j,\,m'_j}$,
$j=\overline{1,\,3}$ are used in obtaining the matrix
$\underline{\mathcal{G}}$ of the six-dimensional Green's function
operator $\hat{\mathcal{G}}$. It has been shown in \cite{JP3} that,
e.g., $\underline{\mathcal{G}}^{(+)}$ can be represented in the form
of a double integral over the complex variables $\mathcal{E}_1$ and
$\mathcal{E}_2$ along straight-line paths (see Figure~\ref{fig1}),
on which $\mathcal{E}_1$, $\mathcal{E}_2$ are parametrized by
\begin{equation}\label{E13}
    \mathcal{E}_1 = \frac{k^2_{23}}{2}+E_1\,e^{i \varphi}, \quad              
    \mathcal{E}_2 = \frac{k^2_{13}}{2}+E_2\,e^{i \varphi},               
\end{equation}
where $E_1$, $E_2$ are real and $-\pi < \varphi <0$. Namely, we have
\begin{equation}\label{z1z2z3}
 \begin{array}{c}
    \left[\underline{\mathcal{G}}^{(+)}\right]_{\mathfrak{N},\,\mathfrak{N}'}=
    \frac{e^{2 i \varphi}}{(2\pi i)^2}\frac{1}{\mu_{23}\,\mu_{13}}\int \limits _{-\infty}^{\infty} \int
    \limits _{-\infty}^{\infty} dE_1\, dE_2\,
    G^{(+)}_{n_1\,m_1;\; n_1'\, m_1'}\left(t_{23};\;\frac{k_{23}^2}{2}+E_1\,e^{i \varphi}\right)\\[4mm]
    \times G^{(+)}_{n_2\,m_2;\; n_2'\,
    m_2'}\left(t_{13};\;\frac{k_{13}^2}{2}+E_2\,e^{i \varphi}\right)
    \,G^{(+)}_{n_3\,m_3;\; n_3'\,
    m_3'}\left(t_{12};\; \mathcal{E}_3\right).  \\
 \end{array}
\end{equation}
Here $\mathcal{E}_3$ is given by
\begin{equation}\label{y3}
\mathcal{E}_3 =
    \frac{k^2_{12}}{2}-\left(\frac{\mu_{12}}{\mu_{23}}\,E_1+
    \frac{\mu_{12}}{\mu_{13}}\,E_2\right)e^{i \varphi}, \quad \left|\arg\left(\mathcal{E}_3\right)
\right|< \pi,
\end{equation}
as follows from (\ref{Cj}) and (\ref{Egma}).

Aside from the replacement $\hat{\mathcal{V}} \rightarrow
\hat{\mathcal{V}}^{\mathcal{N}}$, we make an approximation, which
consists in ignoring the correct boundary conditions in two-body
asymptotic domains (the asymptotic behavior of the C3 wave function
in the neighborhoods of the regions $\Omega_j$, $j=1,\, 2,\, 3$ has
been obtained in \cite{BL11}).

Further approximations are introduced in the treatment of the
potential operator.

\subsection*{The potential operator}

As an example of a three-body Coulomb system above the threshold for
total break-up, we consider a final state for double ionization of
helium. Thus, the potential $\hat{\mathcal{V}}$ is given by
(\ref{V}) and (\ref{D1}). In order to calculate the matrix of
$\hat{\mathcal{V}}$ in the basis (\ref{B6}), we need to express
(\ref{D1}) in terms of the parabolic coordinates. From (\ref{pc}),
it is easy to obtain
\begin{eqnarray}
\label{a1}
  \hat{\bf r}_{13}\cdot\hat{\bf r}_{12} &=& \frac{\left(\xi_2+\eta_2 \right)^2
  +\left(\xi_3+\eta_3 \right)^2-\left(\xi_1+\eta_1 \right)^2}
  {2\left(\xi_2+\eta_2 \right)\left(\xi_3+\eta_3 \right)}, \\
\label{a2}
  \hat{\bf r}_{23}\cdot\hat{\bf r}_{12} &=& \frac{\left(\xi_2+\eta_2 \right)^2
  -\left(\xi_3+\eta_3 \right)^2-\left(\xi_1+\eta_1 \right)^2}
  {2\left(\xi_1+\eta_1 \right)\left(\xi_3+\eta_3 \right)}.
\end{eqnarray}
Whereas, evaluation of matrix elements of scalar products $\hat{\bf
r}_{i\,j}\cdot\hat{\bf k}_{l\,s}$ with $\left\{i,\,j\right\}\neq
\left\{l,\,s\right\}$ in the general case requires the inversion of
the transformation (\ref{pc}). This (numerical) procedure is reduced
to finding roots of a quartic polynomial (see, e. g.,
\cite{Gasaneo97}). Thus, it might appear that the corresponding
Cartesian coordinates ${\bf r}_{ls}$ are complex. To simplify
matters, we take ${\bf k}_{13}={\bf k}$ and ${\bf k}_{23}=-{\bf k}$
and therefore ${\bf k}_{12}={\bf k}$. In this case we have from
(\ref{pc})
\begin{equation}
 \label{a123}
  \begin{array}{c}
  \hat{\bf r}_{13}\cdot \hat{\bf k}_{12}= \frac{1}{r_{13}}\,\hat{\bf k}_{13}\cdot{\bf r}_{13}
  = \frac{\xi_2-\eta_2}{\xi_2+\eta_2}, \\
  \hat{\bf r}_{23}\cdot \hat{\bf k}_{12}= -\frac{1}{r_{23}}\,\hat{\bf
k}_{23}\cdot{\bf r}_{23}=- \frac{\xi_1-\eta_1}{\xi_1+\eta_1}, \\
  \hat{\bf r}_{12}\cdot \hat{\bf k}_{13}=-\hat{\bf r}_{12}\cdot \hat{\bf k}_{23}= \frac{1}{r_{12}}\,\hat{\bf
k}_{12}\cdot{\bf r}_{12}= \frac{\xi_3-\eta_3}{\xi_3+\eta_3}.\\
  \end{array}
\end{equation}

Thus, the matrix of the potential operator $\hat{\mathcal{V}}$ can
be constructed in closed form without the need for numerical
integration. However, it should be noted that in the domains of
integration $\overline{\Omega}_1: \;
\xi_1+\eta_1>\xi_2+\eta_2+\xi_3+\eta_3$ and
$\overline{\Omega}_2:\;\xi_2+\eta_2>\xi_1+\eta_1+\xi_3+\eta_3$ the
triangle inequality is violated, so that $\hat{\bf
r}_{13}\cdot\hat{\bf r}_{12} < -1$ and $\hat{\bf
r}_{23}\cdot\hat{\bf r}_{12} > 1$ within these domains. Hence the
terms $\hat{\bf r}_{13}\cdot\hat{\bf r}_{12}$ and $\hat{\bf
r}_{23}\cdot\hat{\bf r}_{12}$ grow without bound in
$\overline{\Omega}_1$ and $\overline{\Omega}_2$, respectively.
Obviously, such behavior is inconsistent with compactness of the
equation. Actually, our calculations have shown that the use of
$\hat{\bf r}_{13}\cdot\hat{\bf r}_{12}$ (\ref{a1}) and $\hat{\bf
r}_{23}\cdot\hat{\bf r}_{12}$ (\ref{a2}) in the potential operator
leads to divergence. To avoid this problem, adequate analytical
continuations of these scalar products into the regions
$\overline{\Omega}_1$ and $\overline{\Omega}_2$ should be performed.
Here we approximate $\hat{\bf r}_{13}\cdot\hat{\bf r}_{12}$ and
$\hat{\bf r}_{23}\cdot\hat{\bf r}_{12}$ by their projections in the
direction $\hat{\bf k}$:
\begin{eqnarray}
\label{a11}
  \hat{\bf r}_{13}\cdot\hat{\bf r}_{12} & \cong & \left(\hat{\bf r}_{13}\cdot\hat{\bf k}\right)
  \left(\hat{\bf r}_{12}\cdot\hat{\bf k}\right)
  = \frac{\left(\xi_2-\eta_2 \right)\left(\xi_3-\eta_3 \right)}
  {\left(\xi_2+\eta_2 \right)\left(\xi_3+\eta_3 \right)}, \\
\label{a21}
  \hat{\bf r}_{23}\cdot\hat{\bf r}_{12} & \cong & \left(\hat{\bf r}_{23}\cdot\hat{\bf k}\right)
  \left(\hat{\bf r}_{12}\cdot\hat{\bf k}\right)
  = -\frac{\left(\xi_1-\eta_1 \right)\left(\xi_3-\eta_3 \right)}
  {\left(\xi_1+\eta_1 \right)\left(\xi_3+\eta_3 \right)}.
\end{eqnarray}
Note that at least the absolute values of the terms on the
right-hand side of (\ref{a11}) and (\ref{a21}), as well as of
(\ref{a123}), are bounded by 1. Thus we obtain the approximate
formula
\begin{equation}\label{V1}
 \begin{array}{c}
\hat{\mathcal{V}}= 2\left(\xi_2+\eta_2\right)\,\left(-\xi_1
\frac{\partial}{\partial \xi_1}\,\xi_3 \frac{\partial}{\partial
\xi_3}+ \xi_1\frac{\partial}{\partial
\xi_1}\,\eta_3\frac{\partial}{\partial \eta_3}
+\eta_1\frac{\partial}{\partial \eta_1}\,\xi_3\frac{\partial}
{\partial \xi_3}-\eta_1\frac{\partial}{\partial \eta_1}\,\eta_3\frac{\partial}{\partial \eta_3}\right)\\[3mm]
+2\left(\xi_1+\eta_1\right)\,\left(-\xi_2 \frac{\partial}{\partial
\xi_2}\,\xi_3 \frac{\partial}{\partial \xi_3}+
\xi_2\frac{\partial}{\partial \xi_2}\,\eta_3\frac{\partial}{\partial
\eta_3} +\eta_2\frac{\partial}{\partial
\eta_2}\,\xi_3\frac{\partial}
{\partial \xi_3}-\eta_2\frac{\partial}{\partial \eta_2}\,\eta_3\frac{\partial}{\partial \eta_3}\right).\\
 \end{array}
\end{equation}
Further, we take into account only the mixed derivatives
$\partial^2/\partial \xi_1\partial \xi_3$ and $\partial^2/\partial
\xi_2\partial \xi_3$. Thus, in our calculations, we use the
potential
\begin{equation}\label{Va}
\hat{{V}}= -2\left[\left(\xi_2+\eta_2\right)\,\xi_1
\frac{\partial}{\partial \xi_1}\,\xi_3 \frac{\partial}{\partial
\xi_3} +\left(\xi_1+\eta_1\right)\,\xi_2 \frac{\partial}{\partial
\xi_2}\,\xi_3 \frac{\partial}{\partial \xi_3}\right],
\end{equation}
which corresponds to the outgoing approximation \cite{Gasaneo97}.

\subsection*{The inhomogeneity}
The expansion coefficients
$\left[\underline{a}^{(0)}\right]_{\mathfrak{N}}$ of the
inhomogeneity $\Psi_{C3}$ (\ref{PsiC3}) is written in terms of
polynomials $p_n$ \cite{JP2}:
\begin{equation}\label{a0n}
 \left[\underline{a}^{(0)}\right]_{\mathfrak{N}}=\prod \limits
 _{j=1}^3 \frac{2}{b_j}\left[2\left(b_j+i k_{ls} \right)
 \right]^{-it_{ls}}(-1)^{m_j}p_{n_j}\left(t_{ls}+\frac{i}{2};\; \frac{b_j-i k_{ls}}{b_j+i
 k_{ls}}\right),
\end{equation}
\begin{equation}\label{pn}
  p_n(\tau;\; \zeta)=\frac{(-1)^n}{n!}\frac{\Gamma\left(n+\frac{1}{2}-i\tau \right) }
  {\Gamma\left(\frac{1}{2}-i\tau \right) }\;
  {_2F_1\left(-n,\,\frac{1}{2}+i\tau;\; -n+\frac{1}{2}+i\tau;\; \zeta
  \right)}.
\end{equation}
The constant factor
\begin{equation}\label{a00}
 \left[\underline{a}^{(0)}\right]_{0}=\prod \limits
 _{j=1}^3 \frac{2}{b_j}\left[2\left(b_j+i k_{ls} \right)
 \right]^{-it_{ls}}
\end{equation}
is omitted below for simplicity, so that we consider the ``reduced''
coefficients
\begin{equation}\label{aredn}
 \left[\overline{\underline{a}}\right]_{\mathfrak{N}} \equiv
 \left[\underline{a}\right]_{\mathfrak{N}}/\left[\underline{a}^{(0)}\right]_{0}.
\end{equation}

\section{Results for the $(e^-,\, e^-,\, {\mbox{He}^{++}})$ system}
Let us consider the case of a back-to-back electron emission with
equal energy sharing. We put $k_{ls}=k=1.5$ and choose the values of
the exponents $b_j$ in the basis to be equal to the wave number, i.
e., $b_j=b=1.5$, $j=1,\,2,\,3$.

We use the single basis function $\psi_0$ (\ref{B1}) for the
parabolic coordinates $\eta_j$ and up to sixteen functions for each
of the three coordinates $\xi_j$ in the potential operator expansion
(\ref{VN}). Thus, we put $M_1=M_2=M_3=1$ and examine the convergence
behavior of the first expansion coefficient
$\left[\overline{\underline{a}}\right]_{\mathfrak{N}}$ as the number
$N=N_1=N_2=N_3$ of the basis functions
$\psi_{n_j}\left(\xi_j\right)$, $j=1,\,2,\,3$ is increased.

\begin{figure*}[ht]
\centerline{\psfig{figure=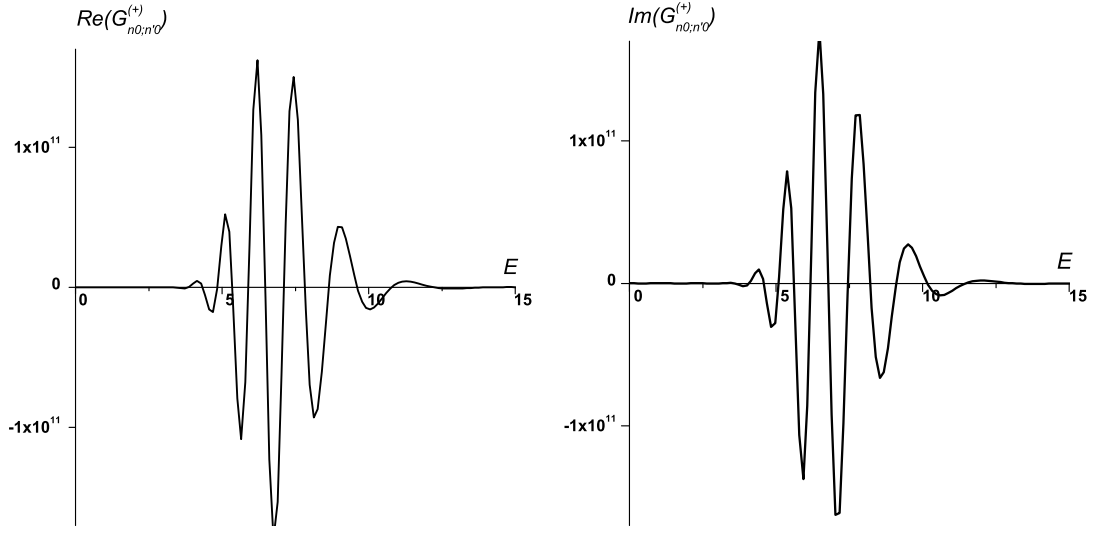,width=1\textwidth}}
\caption{The matrix element
$G^{(+)}_{n,\,0;\;n',\,0}\left(-\frac{2}{k},\, \mathcal{E}\right)$
(\ref{Gnmn_mp}) with $n=n'=20$ and $m=m'=0$ on the contour
$\mathcal{C}$ in Figure~\ref{fig1} for $\varphi=-\frac{\pi}{2}$.}
\label{fig2}
\end{figure*}

\begin{figure*}[ht]
\centerline{\psfig{figure=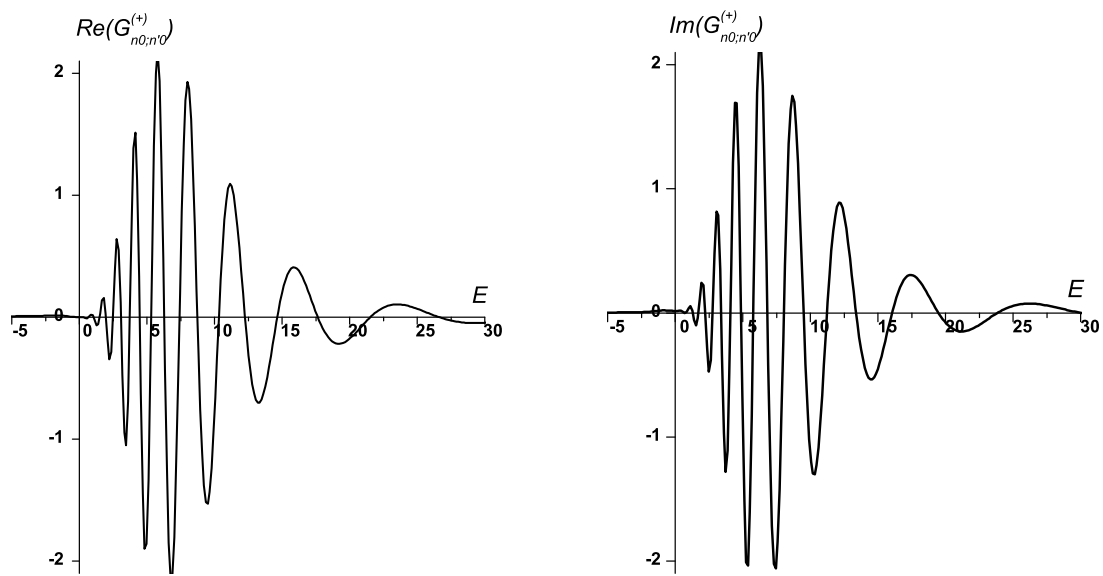,width=1\textwidth}}
\caption{The same as in Figure~\ref{fig2} but for
$\varphi=-\frac{\pi}{6}$.} \label{fig3}
\end{figure*}

The only practical limitation on the total number $\mathcal{N}=N^3$
of the basis functions arises from the difficulty of computing the
matrix elements (\ref{z1z2z3}) of the three-body Coulomb Green's
function operator with sufficient numerical accuracy. Actually, the
integrand includes oscillatory functions whose amplitude grows very
rapidly as the indices of the basis functions increase. As an
example, Figure~\ref{fig2} shows the matrix element
$G^{(+)}_{n,\,0;\;n',\,0}\left(-\frac{2}{k},\, \mathcal{E}\right)$
with $n=n'=20$ and $\mathcal{E}=\frac{k^2}{2}+E e^{i \varphi}$,
$\varphi=-\frac{\pi}{2}$. A comparison with Figure~\ref{fig3} shows
that a relatively small change in the value of the angle $\varphi$
can produce a large change in the amplitude. Note that in order to
evaluate the matrix elements $G^{(+)}_{n,\,0;\;n',\,0}$
(\ref{Gnmn_mp}) of the two-dimensional Green' function we resort to
quadruple length arithmetic.

The coefficients
$\left[\overline{\underline{a}}\right]_{\mathfrak{N}}$ are found to
exhibit oscillations whose amplitude grows as $N$ increases. A
simple way to damp the oscillations, is to multiply each matrix
element $\left[\underline{\mathcal{V}}
\right]_{\mathfrak{N},\,\mathfrak{N}'}$ by the Lanczos smoothing
factors
\begin{equation}\label{SF}
    \sigma_n^N=\frac{1-\exp\left\{-\left[\alpha(n-N)/N
    \right]^2\right\}}{1-\exp(-\alpha^2)},
\end{equation}
which attenuate $\left[\underline{\mathcal{V}}
\right]_{\mathfrak{N},\,\mathfrak{N}'}$ with large indices $n_j,\,
n'_j$. Thus, $\left[\underline{\mathcal{V}}
\right]_{\mathfrak{N},\,\mathfrak{N}'}$ in the expansion (\ref{VN})
are replaced by
\begin{equation}\label{Vnnp}
\left[\underline{\mathcal{V}}
\right]_{\mathfrak{N},\,\mathfrak{N}'}\,\prod \limits
_{j=1}^{3}\sigma_{n_j}^N\,\sigma_{n'_j}^N.
\end{equation}
The optimal value for the parameter $\alpha$ in (\ref{SF}) is
$\alpha \approx 3$. The convergence behavior of the coefficient
$\left[\overline{\underline{a}}\right]_0$ ($n_j=m_j=0$,
$j=1,\,2,\,3$) as the number $N$ is increased is presented in
Figure~4. Note that such a simple remedy allows one to achieve
convergence in the framework of the algebraic approach to
two-particle scattering problems provided that the long-range part
of the Hamiltonian is included into the ``free'' Green's function.
The results for the first few coefficients
$\left[\underline{\overline{a}}\right]_{\mathfrak{N}}$ are shown in
Table~\ref{T1}. In this calculation we obtained adequate convergence
by including $N\approx 15$ basis functions (\ref{B1}) for the
coordinates $\xi_j$, $j=1,\, 2,\, 3$.

\section{Conclusion}

The three-body continuum has been treated in terms of the
generalized parabolic coordinates. A back-to-back electron emission
from helium atom has been chosen as an example. The
Lippmann-Schwinger type equation for the continuum-state wave
function has been solved numerically within the framework of the
parabolic Sturmians approach.

The potential in the basic integral equation is represented by the
non-orthogonal part $\hat{D}_1$ of the kinetic energy operator.
$\hat{D}_1$ treated in terms of the Cartesian coordinates ${\bf
r}_{13}$ and ${\bf r}_{23}$ is a bounded operator. However, the
change of variables ${\bf r}_{13},\,{\bf r}_{23} \rightarrow
\xi_j,\, \eta_j$, $j=1,\,2,\,3$ (\ref{pc}) transforms this operator
into an unbounded one. Actually, the kernel of the operator
$\hat{D}_1$ grows without bound in the regions $\overline{\Omega}_1$
and $\overline{\Omega}_2$. On the other hand, the triangle
inequality for the vectors ${\bf r}_{13}$, ${\bf r}_{23}$ is
violated in these regions and therefore there does not exist a
region in the real Cartesian coordinate system which corresponds to
$\overline{\Omega}_1$ or $\overline{\Omega}_2$. Hence, this problem
is cured by defining the kernel of $\hat{D}_1$ in the regions
$\overline{\Omega}_1$ and $\overline{\Omega}_2$ in an appropriate
way (e. g., by setting the kernel equal to zero in these regions).
For this purpose, we have approximated $\hat{\bf r}_{13}\cdot
\hat{\bf r}_{12}$ and $\hat{\bf r}_{23}\cdot \hat{\bf r}_{12}$ which
appear in the expression for $\hat{D}_1$ by products of the
projections of the vectors $\hat{\bf r}_{ls}$ in the direction
$\hat{\bf k}_{12}$.

Besides, we have used the so-called outgoing approximation which
assumes that the sought-for solution depends only on $\xi_j$,
$j=1,\,2,\,3$. Then the resulting integral equation compactness has
been tested numerically by utilizing the parabolic Sturmian basis
representation of the potential. In particular, convergence of the
expansion coefficients of the solution has been obtained as the size
of the basis is enlarged.

\section*{Acknowledgments}
We are thankful to the Computer Center, Far Eastern Branch of the
Russian Academy of Science (Khabarovsk, Russia) for generous
rendering of computer resources to our disposal. Additional thanks
are expressed to Dr. V.~Borodulin for his kind hospitality and help.


\newpage

\newpage

\newpage

\newpage
\begin{figure*}[ht]
\centerline{\psfig{figure=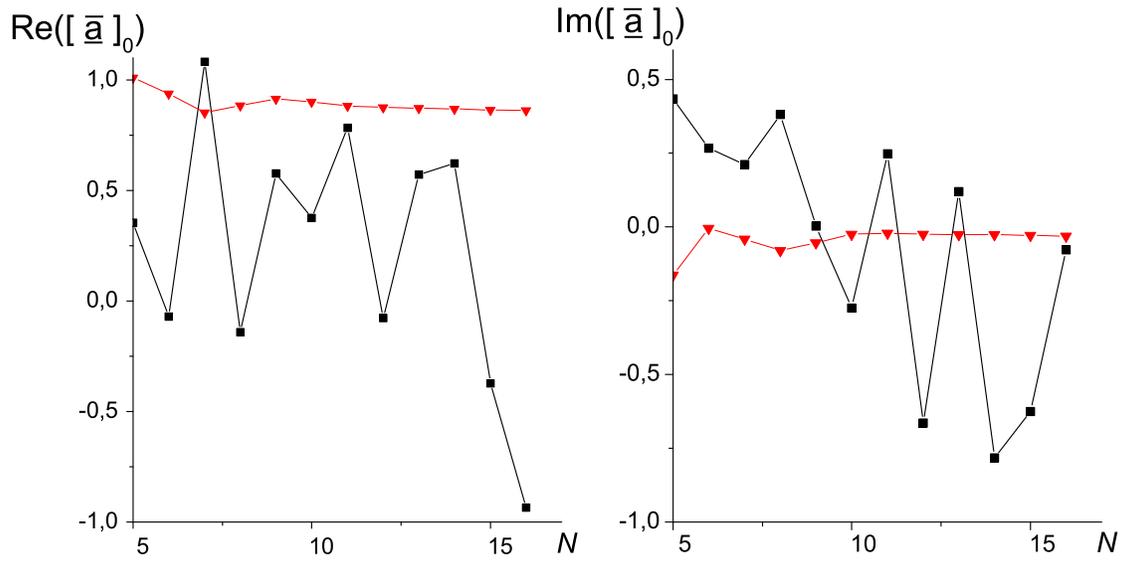,width=1\textwidth}}
\caption{The convergence behavior of
$\left[\overline{\underline{a}}\right]_0$ as the number $N$ of basis
functions (\ref{B1}) for each of the coordinates $\xi_1,\, \xi_2,\,
\xi_3$ is increased. The coefficient values obtained without and
with smoothing factors are denoted by squares and triangles,
respectively.} \label{fig4}
\end{figure*}

\newpage

\begin{table}
 \caption{Convergence of the first few ``reduced'' coefficients $\left[\underline{\overline{a}}\right]_{\mathfrak{N}}$.
 $N$ is the number of the basis functions (\ref{B1}) ($N=N_1=N_2=N_3$) in the expansion (\ref{VN}).}\label{T1}
\begin{ruledtabular}
{\small
\begin{tabular}{c|c|c|c}
 $N$ & $n_1=n_2=n_3=0$ & $n_1=1,\,n_2=n_3=0$ & $n_1=n_2=0,\, n_3=1$ \\
\hline
  0 &  $1.00000000+i 0.00000000$ & $0.33332667-i 1.33332667$& $-1.33333333+i 0.33333333$ \\
\hline
  5 &   $1.00869682-i 0.16338707$ & $0.06484035-i 1.03904748$ & $-0.89298931+i 0.33045941$ \\
  6 &   $0.93751779-i 0.00481271$ & $0.21244616-i 1.16523626$ & $-1.09329190+i 0.31413236$ \\
  7 &   $0.85258743-i 0.04143155$ & $0.28687445-i 1.10607051$ & $-1.09100360+i 0.21980636$ \\
  8 &   $0.88418841-i 0.08046765$ & $0.25361563-i 1.06748525$ & $-1.04328109+i 0.23545010$ \\
  9 &   $0.91352864-i 0.05297516$ & $0.22333705-i 1.09569062$ & $-1.06280041+i 0.27685677$ \\
  10 &  $0.90101510-i 0.02409631$ & $0.23719582-i 1.12705918$ & $-1.10091006+i 0.27238506$ \\
  11 &  $0.88184853-i 0.02083012$ & $0.25897978-i 1.13142201$ & $-1.11043846+i 0.25184571$ \\
  12 &  $0.87486976-i 0.02519219$ & $0.26812774-i 1.12658516$ & $-1.10795951+i 0.24433839$ \\
  13 &  $0.87207818-i 0.02578117$ & $0.27172914-i 1.12525415$ & $-1.11016454+i 0.24216317$ \\
  14 &  $0.86790716-i 0.02592191$ & $0.27617970-i 1.12505442$ & $-1.11300765+i 0.23690466$ \\
  15 &  $0.86415304-i 0.02795984$ & $0.28050647-i 1.12351749$ & $-1.11247348+i 0.23154456$ \\
  16 &  $0.86200713-i 0.03060950$ & $0.28381235-i 1.12130593$ & $-1.11061131+i 0.22824086$ \\
\end{tabular}}
\end{ruledtabular}
\end{table}


\begin{thebibliography}{99}
\bibitem{CCC1}
 I.~Bray and A.~T.~Steblovics, Phys. Rev. Lett., {\bf 70}, 746 (1993).

\bibitem{CCC2}
 I.~Bray, D.~V.~Fursa, A.~S.~Kheifets, and A.~T.~Steblovics,
J. Phys. B,  {\bf 35}, R117 (2002).

\bibitem{TPF1}
 Z.~Papp, Phys. Rev. C, {\bf 55}, 1080 (1997).

\bibitem{TPF2}
 Z.~Papp, C-.~Y.~Hu , Z.~T.~Hlousek, B.~K\'{o}nya, and S.~L.~Yakovlev,
 Phys. Rev. A, {\bf 63}, 062721 (2001).

\bibitem{JM1}
The J-Matrix Method: Developments and Applications, Ed. by
A.~D.~Alhaidari, E.~J.~Heller, H.~A.~Yamani, and M.~S.~Abdelmonem
(Springer Sci., Business Media, 2008).

\bibitem{JM2}
S.~A.~Zaytsev, V.~A.~Knyr,  Yu.~V.~Popov, A.~Lahmam-Bennani, Phys.
Rev. A, {\bf 76}, 022718 (2007).

\bibitem{JM3}
M.~Silenou Mengoue, M.~G.~Kwato Njock, B.~Piraux, Yu.~V.~Popov, and
S.~A.~Zaytsev, Phys. Rev. A, {\bf 83}, 052708 (2011).

\bibitem{Sturm1}
A.~L.~Frapiccini, J.~M.~Randazzo, G.~Gasaneo, and F.~D.~Colavecchia,
J. Phys. B,  {\bf 43}, 101001 (2010).



\bibitem{Kadyrov8}
 A.~S.~Kadyrov, I.~Bray, A.~M.~Mukhamedzhanov, and A.~T.~Steblovics,
 Phys. Rev. Lett., {\bf 101}, 230405 (2008).


\bibitem{JP1}
 S.~A.~Zaytsev, J. Phys. A,  {\bf 41} 265204 (2008).

\bibitem{JP2}
 S.~A.~Zaytsev, J. Phys. A,  {\bf 42} 015202 (2009).

\bibitem{JP3}
 S.~A.~Zaytsev, J. Phys. A,  {\bf 43} 385208 (2010).


\bibitem{Klar}
 H.~Klar, Z. Phys. D, {\bf 16}, 231 (1990).

\bibitem{Redmond}
L.~Rosenberg, Phys. Rev. D, {\bf 8}, 1833 (1973).

\bibitem{C31}
 Dz.~Belkic, J. Phys. B, {\bf 11}, 3529 (1978).

\bibitem{C32}
 C.~R.~Garibotti, J.~E.~Miraglia, Phys. Rev. A, {\bf 21}, 572 (1980).

\bibitem{C33}
 M.~Brauner, J.~S.~Briggs, H.~Klar, J. Phys. B, {\bf 22}, 2265 (1989).

\bibitem{Jones2003}
 S.~Jones and D.~H.~Madison, Phys. Rev. Lett., {\bf 91}, 073201 (2003).

\bibitem{Ancarani2004}
 L.~U.~Ancarani, T.~Montagnese, and C.~Dal~Capello, Phys. Rev. A, {\bf 70}, 012711 (2004).

\bibitem{Ch10}
 O.~Chuluunbaatar, H.~Bachau, Yu.~V.~Popov, B.~Piraux,
 K.~Stefa\'{n}ska, Phys. Rev. A, {\bf 81}, 063424 (2010).

\bibitem{Gasaneo97}
 P.~A.~Macri, J.~E.~Miraglia, C.~R.~Garibotti, F.~D.~Colavecchia
 and G.~Gasaneo, Phys. Rev. A, {\bf 55}, 3518 (1997).

\bibitem{Ojha1}
 P.~C.~Ojha, J. Math. Phys. {\bf 28}, 392 (1987).


\bibitem{Canuto}
C.~Canuto, A.~Quarteroni, M.~Y.~Hussaini, T.~A.~Zang, Spectral
Methods. Fundamentals in Single Domains (Springer-Verlag, Berlin,
Heidelberg, 2006).

\bibitem{Stone}
M.~Stone, Mathematics for Physics I (Pimander-Casaubon, Alexandria,
Florence, London, 2002).


\bibitem{PSE}
 J.~R\'{e}vai, M.~Sotona, and J.~\v{Z}ofka, J. Phys. G, {\bf 11}, 745 (1985).

\bibitem{Papp0}
 B.~K\'{o}nya, G.~L\'{e}vai, and Z.~Papp, Phys. Rev. C, {\bf 61} 034302 (2000).

\bibitem{Berakdar}
J.~Berakdar, Phys. Rev. A, {\bf 53}, 3214 (1996).

\bibitem{Swainson}
 R.~A.~Swainson, G.~W.~Drake, J. Phys. A,  {\bf 24} 95 (1991).

\bibitem{Abramowitz}
 M.~Abramowitz and I.~A.~Stegun, Handbook of mathematical functions (New York: Dover),
 1970.

\bibitem{FMB}
 L.~D.~Faddeev and S.~P.~Merkuriev, Quantum Scattering Theory for
 Several Particle Systems (Kluwer Academic Publishers, Dordrecht,
 1993).


\bibitem{BL11}
 V.~S.~Buslaev, S.~B.~Levin, 2011, arXiv:1104.3358v1 [math-ph].



\end{thebibliography}
\end{document}